\documentclass[useAMS,usenatbib]{mn2e}
\usepackage{epsfig,natbibmnfix}
\usepackage{amsmath, amssymb, aas_macros}
\usepackage{graphicx, subfig}

\def\simeq{
\mathrel{\raise.3ex\hbox{$\sim$}\mkern-14mu\lower0.4ex\hbox{$-$}}
}

\def\msun{{\rm M_{\odot}}}

\def\be{\begin{equation}}
\def\ee{\end{equation}}

\def\le{{L_{\rm Edd}}}
\def\msun{{\rm M_{\odot}}}

\def\del#1{{}}

\title{The $M - \sigma$ relation in different environments}

\author[K. Zubovas, A.R. King] {\parbox{18cm}{K. Zubovas$^{1}$,
    A.R. King$^{1}$}\\ $^1$Theoretical Astrophysics Group, University
  of Leicester, Leicester LE1 7RH}

\begin{document}

\maketitle

\begin{abstract}

Galaxies become red and dead when the central supermassive black hole (SMBH)
becomes massive enough to drive an outflow beyond the virial radius of the
halo. We show that this final SMBH mass is larger than the final SMBH mass in
the bulge of a spiral galaxy by up to an order of magnitude. The $M - \sigma$
relations in the two galaxy types are almost parallel ($M \propto
\sigma^{4+\beta}$, with $\beta < 1$) but offset in normalization, with the
extra SMBH mass supplied by the major merger transforming the galaxy into an
elliptical, or by mass gain in a galaxy cluster.  This agrees with recent
findings that SMBH in two Brightest Cluster Galaxies are $\sim 10\times$ the
expected $M-\sigma$ mass. We show that these results do not strongly depend on
the assumed profile of the dark matter halo, so analytic estimates found for
an isothermal potential are approximately valid in all realistic cases.

Our results imply that there are in practice actually {\it three} $M - \sigma$
relations, corresponding to spiral galaxies with evolved bulges, field
elliptical galaxies and cluster centre elliptical galaxies. A fourth relation,
corresponding to cluster spiral galaxies, is also possible, but such galaxies
are expected to be rare. All these relations have the form $M_{\rm BH} =
C_n\sigma^4$, with only slight difference in slope between field and cluster
galaxies, but with slightly different coefficients $C_n$.  Conflating data
from galaxies of different types and fitting a single relation to them tends
to produce a higher power of $\sigma$.

\end{abstract}


\renewcommand{\thefootnote}{\fnsymbol{footnote}}
\footnotetext[1]{E-mail: {\tt kastytis.zubovas@le.ac.uk }}

\section{Introduction}

In the past decade, observations of large numbers of galaxies have revealed
important correlations between the properties of galaxy spheroids and
the supermassive black holes (SMBHs) they host. Two of these relations
are particularly interesting: the black hole mass -- host spheroid velocity
dispersion ($M - \sigma$) relation \citep{Ferrarese2006ApJ} and the black hole
- bulge mass ($M - M_{\rm b}$) relation \citep{Haering2004ApJ}. Together, they
strongly imply that SMBHs coevolve with their hosts, and probably affect
each other's evolution through some form of feedback.

There have been numerous attempts to explain these relations, both
analytically \citep{Silk1998A&A, King2003ApJ, King2005ApJ, Murray2005ApJ,
  Power2011MNRAS} and numerically \citep[e.g.][]{DiMatteo2005Natur,
  Booth2009MNRAS}. The AGN wind feedback model \citep[and references
  therein]{King2003ApJ, King2005ApJ, King2010MNRASa, Zubovas2012ApJ} is
particularly promising, as it explains not only the observed correlations, but
also other observable properties of galactic winds
\citep[e.g.][]{Pounds2003MNRASb, Pounds2003MNRASa,Pounds2011MNRAS,
  Tombesi2010A&A, Tombesi2010ApJ} and outflows\footnote{For clarity, we
  distinguish between the black hole {\it wind}, the gas coming directly from
  the accretion disc around the SMBH, and the {\it outflow}, the outward
  movement of swept--up interstellar gas from the host galaxy. These two
  components undergo respectively reverse and forward shocks on each side of
  the contact discontinuity separating them. See the Figure in
  \citet{Zubovas2012ApJ} for more details} \citep[e.g.][]{Rupke2011ApJ,
  Sturm2011ApJ}. In this picture, the wind and outflow shocks initially occur
fairly close to the SMBH, and are efficiently Compton--cooled. The outflow
retains only the ram pressure $\simeq L_{\rm Edd}/c$ of the black hole wind
and has kinetic energy lower by a factor $\sigma/c \sim 10^{-3}$. Only when
the SMBH mass approaches the critical mass $M_{\sigma} \simeq 3.7 \times
  10^8 \; \msun$ (see equation \ref{eq:msigma} below) do the shocks move
further away. The resulting geometrical dilution of the quasar radiation field
now makes Compton cooling inefficient. The shocks are no longer cooled, and
the outflows become energy--driven.

In two recent papers \citep{King2011MNRAS, Zubovas2012ApJ} we investigated the
properties of large (kiloparsec)--scale outflows in galaxies.  These must be
energy--driven, i.e. the outflow kinetic energy rate $\dot{E}_{\rm out} =
\dot{M}_{\rm out} v_{\rm out}^2 / 2 \simeq 0.05 L_{\rm Edd}$, where $L_{\rm
  Edd}$ is the Eddington luminosity of the driving quasar.  We showed that
energy--driven outflows are able to clear galaxies of gas, turning them into
red--and--dead spheroids.

However, in these papers, we only briefly touched upon the question of how
much the SMBH mass grows as it expels the outflow from the galaxy, and this
remains uncertain. There are two discrepant claims: \citet{Power2011MNRAS}
used an energy argument to conclude that the SMBH only needs to grow its mass
by $\sim40\%$ to expel the surrounding gas past the virial radius, while in
\citet{King2011MNRAS} we found that an outburst duration of $\sim 10^8$~yr is
required, which would lead to the SMBH mass growing by about an order of
magnitude.

We return to this problem in this paper. We also consider the difference
between evolved spiral and elliptical galaxies in more general terms, as well
as the dependence of the outflow properties on the galaxy environment. We find
only a small uncertainty in our results stemming from the assumption of an
idealized background potential (Section \ref{sec:propagation}). By considering
the distribution of energy in various components of the outflow we resolve the
discrepancy between the two claims (Section \ref{sec:outflows}). We conclude
that SMBHs in elliptical galaxies should grow to a mass a few times higher
than in spirals for a given value of $\sigma$. We consider the effects of gas
depletion and replenishment in a galaxy and find that galaxies in cluster
environments should have slightly more massive SMBHs than their counterparts
in the field (Section \ref{sec:clusters}). We summarize and discuss our
results in Section \ref{sec:discuss}.

\section{Outflows in a General Potential} \label{sec:propagation}

In previous work on the wind feedback model, we adopted an isothermal
background potential, which has the useful property that the
  weight of the gas is independent of radius. To check the
  validity of this assumption we now consider what happens to an
energy--driven (large--scale) flow in a more realistic halo. As an
example we consider an NFW halo \citep{Navarro1997ApJ}. We note that
\citet{McQuillin2012arXiv} have considered this problem assuming
{\it momentum--driven} flows for all radii (also for Hernquist and 
Dehnen--McLaughlin potentials) and shown that the differences from 
the simple isothermal treatment are fairly small.

One particular difference between isothermal and NFW halos that is
  important for us is the varying velocity dispersion. In general, all
  non--isothermal halos have varying $\sigma$, usually  with $\sigma$
increasing from the centre out to some radius $r_{\rm peak}$ and decreasing
beyond it. As the weight of the gas swept up by the outflow depends on
  the value of $\sigma$ through the equation governing its density, it is no
  longer independent of radius, as is the case for an isothermal
  distribution. Specifically, the weight of the gas increases as the outflow
  moves from low radii toward $r_{\rm peak}$ and decreases afterwards. This
means that an SMBH is able to inflate a rather large (but still with $R <
  R_{\rm peak}$) outflow bubble even though its mass may still be too small
to completely drive the gas out of a galaxy. On the other hand, once the
outflow passes $R_{\rm peak}$, it becomes easier to expel the gas.

In the present calculations, we consider an NFW halo with a concentration
parameter \citep{Navarro1997ApJ} $c = 10$, scale radius $R_{\rm s} = 25$~kpc
(so that $R_{\rm V} = 250$~kpc) and virial mass $M_{\rm V} = 3.3 \times
10^{12} \msun$. This gives $\sigma_{\rm peak} \simeq 200$~km s$^{-1}$ at
$R_{\rm peak} \simeq 2.16 R_{\rm s} \simeq 54$~kpc. We take the initial SMBH
mass to be $M_0 = 3.68 \times 10^8 \msun$, equal to the critical SMBH mass
found in previous papers \citep[e.g.][]{King2010MNRASa}. We compare this
  NFW halo with an isothermal halo with $\sigma = 200$~km/s. We do not
  truncate either halo beyond $R_{\rm V}$, but allow them to continue to
  infinity.

\begin{figure*}
  \centering
 \subfloat{\includegraphics[width=0.5\textwidth]{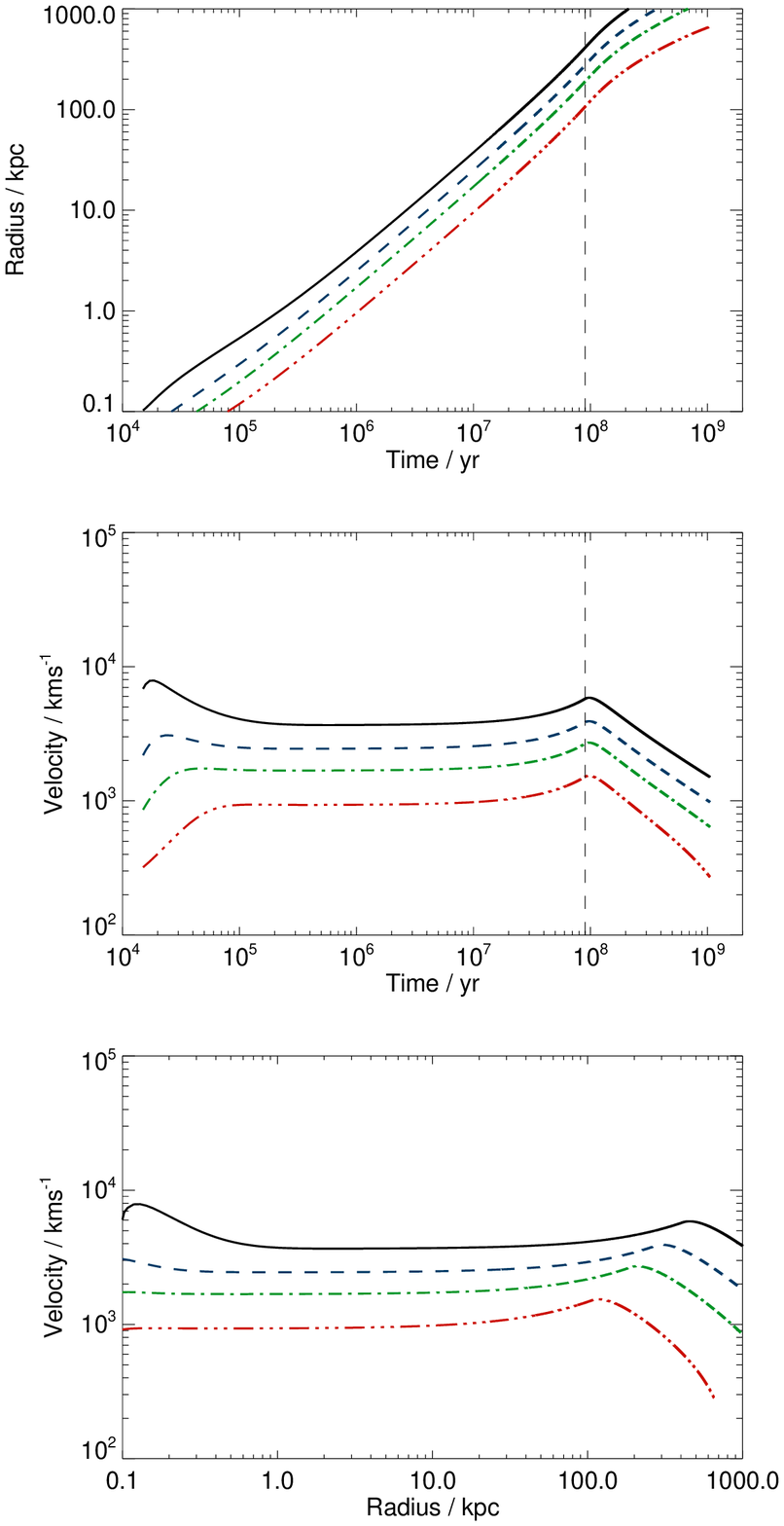}}
  \subfloat{\includegraphics[width=0.5\textwidth]{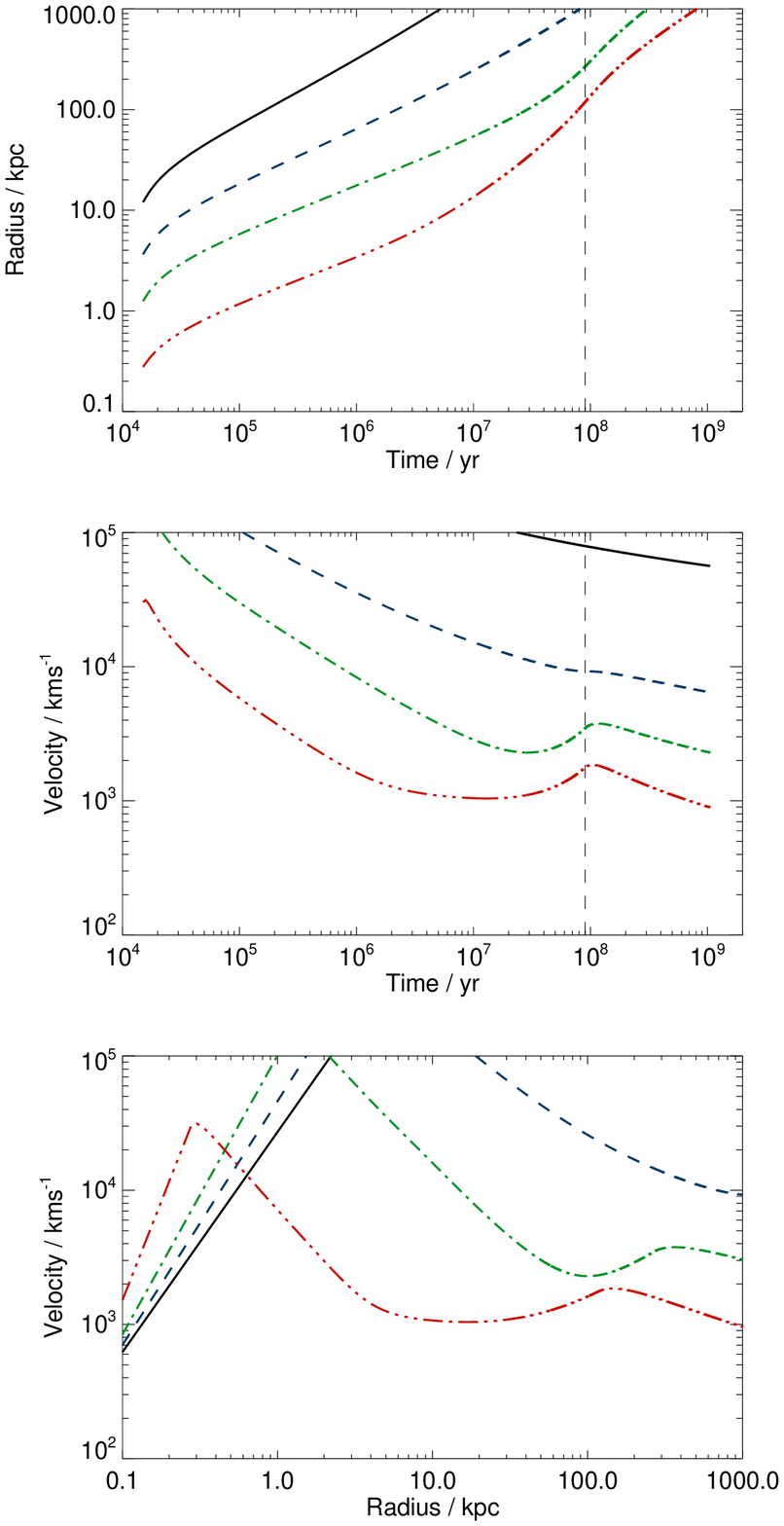}}
  \caption{Propagation of energy-driven outflows in isothermal (left)
    and NFW (right) halos. The four outflows have $f_{\rm g} = 3
    \times 10^{-3}, 0.01, 0.03$ and $0.16$ for the black solid, blue
    dashed, green double--dashed and red triple--dashed curves,
    respectively. The SMBH is active for $90$~Myr (dashed vertical
    line in top and middle panels). As shown in \citet{King2011MNRAS},
    outflow properties do not strongly depend on initial conditions
    (radius and velocity). {\it Top panels}: radius of the contact
    discontinuity as function of time. {\it Middle panels}: outflow
    velocity as function of time. {\it Bottom panels}: outflow
    velocity as function of radius.}
  \label{fig:outflows}
\end{figure*}

The equation of motion for an energy driven--shell in an isothermal potential
was derived by \citet{King2005ApJ} and its solutions investigated in
\citet{King2011MNRAS}. Here we derive a generic equation of motion, valid for
any spherically symmetric mass profile, provided that the gas density traces
the background potential density, i.e. $M_{\rm g}\left(R\right) = f_{\rm g}
M\left(R\right)$ for any $R$, where $f_{\rm g}$ is called the gas fraction. We
start with equations (15) and (16) of \citet{King2005ApJ}:
\be \label{eq:mom} 
f_{\rm g} \frac{{\rm d}}{{\rm
    d}t}\left[M\left(R\right)\dot{R}\right] + \frac{G f_{\rm g}
  M^2\left(R\right)}{R^2} = 4 \pi R^2 P 
\ee 
and 
\be \label{eq:energ}
\frac{{\rm d}}{{\rm d}t}\left[\frac{4 \pi R^3}{3}\times
  \frac{3}{2}P\right] = \frac{\eta}{2}L_{\rm Edd} - P\frac{{\rm
    d}}{{\rm d}t}\left(\frac{4 \pi R^3}{3}\right) - \frac{G f_{\rm g}
  M^2\left(R\right)}{R^2} \dot{R}.  
\ee 
Using eq. (\ref{eq:mom}) to eliminate $P$ from eq. (\ref{eq:energ}) and
expanding the terms with derivatives leads to a generalized equation of motion
(we have dropped the $\left(R\right)$ identifier from each mass for
compactness):
\be \label{eq:eom}
\begin{split}
\frac{\eta}{2 f_{\rm g}} L_{\rm Edd} = & \frac{3}{2}\frac{G M^2 \dot{R}}{R^2} +
\frac{GM\dot{M}}{R} + \frac{3}{2}M\dot{R}\ddot{R} +
\frac{3}{2}\dot{M}\dot{R}^2 + \\ &{MR\dddot{R} \over 2} + \dot{M}R\ddot{R} +
\frac{\ddot{M}R\dot{R}}{2}.
\end{split}
\ee
Here, $\dot{M} = \partial M/ \partial R \times \dot{R}$ and $\ddot{M} =
\partial M/ \partial R \times \ddot{R} + \dot{R}^2 \times \partial^2M /
\partial R^2$. This equation reduces to eq. (18) of \citet{King2005ApJ} in the
case of an isothermal potential.

For an NFW potential the equation cannot be solved analytically, so we
integrate it numerically. This also allows us to follow the growth of the
central black hole mass (which slightly affects $L_{\rm Edd}$ on the left hand
side of (\ref{eq:eom})). Figure \ref{fig:outflows} shows the propagation of
four outflows, with $f_{\rm g} = 3 \times 10^{-3}, 0.01, 0.03$ and $0.16$
(lines from top to bottom in the top panel, respectively), in an isothermal
potential (left) and an NFW potential described above (right). In both cases
the SMBH is radiating and growing at the Eddington rate for $t = 90$~Myr $= 2
t_{\rm Sal}$. Note that our results are valid only where the outflow
  velocity is $v_{\rm out} < v_{\rm w} \simeq 0.1 c$. This condition is
 certainly satisfied when the gas density is high.

We see immediately that in an NFW halo, the outflows reach larger radii than
in the isothermal case. This is expected, as the outflows are pushed past the
radius of highest $\sigma$ and so are able to accelerate to higher velocities
before the central AGN switches off. The outflow velocity is almost constant
during driving in the isothermal case, growing only in response to the growth
of the SMBH ($v \propto M_{\rm BH}^{1/3}$) and decays approximately as $v
\propto t^{-1/2}$ afterwards. In the NFW case, the velocity is not constant,
but the minimum velocity of the densest outflow is reached at $R \simeq R_{\rm
  peak}$ and is similar to $v_{\rm e}$, as might be expected.

From the results of the numerical integration it is clear that the isothermal
potential probably represents a worst case scenario for outflow escape. In
other words, the required AGN activity time and the final SMBH mass obtained
using the isothermal assumption are upper limits. This is 
  reasonable, because the weight of the gas, which is $\propto
  \sigma^4$, is constant for the isothermal halo, whereas it varies and
 is always smaller for the NFW halo, which has $\sigma \leq \sigma_{\rm peak} =
  \sigma_{\rm isotherm}$.

The differences between the isothermal and NFW models are rather small
  for high gas density. For lower gas densities ($f_{\rm g} \ll 0.16$), the
  outflow in an NFW potential reaches much higher velocities and escapes past
  $R_{\rm V}$ much faster. However, since the higher density case is relevant
  for establishing the final SMBH mass \citep{King2003ApJ}, we expect the
  uncertainty due to the shape of the background potential to be modest. This
is reassuring in that earlier treatments of energy--driven outflows
\citep[e.g.][]{King2005ApJ, King2011MNRAS} remain valid. However the question
of the final SMBH mass remains. In the rest of this paper, we analyze this
question in more detail. We use the singular isothermal sphere profile of
  both gas and the background potential. We motivate our choice by the fact
  that analytical calculations are much more difficult, and often impossible,
  for non-isothermal potentials. In addition, using a different profile does
  not necessarily make our results more realistic: the central parts of
  galaxies, which are baryon dominated, are known to have profiles similar to
  isothermal \citep{Naab2007ApJ}, and even their dark matter profiles seem to
  differ from any known analytic profiles.

\section{Properties of large scale outflows} \label{sec:outflows}

The radiation pressure of an active SMBH drives a quasi--spherical wind from
the outskirts of its accretion disc \citep{King2010MNRASa}, which then hits
and shocks against the ambient interstellar medium (ISM). The wind is mildly
relativistic ($v_{\rm w} \sim \eta c \sim 0.1c$, where $\eta$ is the radiative
accretion efficiency), and so its shock temperature is a few times
$10^{11}$~K, much greater than the Compton equilibrium temperature ($T_{\rm C}
\sim 2 \times 10^7$~K). As we remarked above, the shocked wind therefore cools
via the inverse Compton process against the photons of the AGN radiation field
\citep{Ciotti1997ApJ}. If the interaction between the wind and the ISM happens
within a critical cooling radius \citep{King2003ApJ}, the wind loses most of
its original energy and only communicates its ram pressure to the ISM,
creating a momentum--driven outflow \citep{King2010MNRASa}. Outside the
cooling radius, on the other hand, the wind cannot cool efficiently and most
of its energy rate $\dot{E}_{\rm w} \simeq \eta L_{\rm Edd} / 2$ is
communicated to the ISM, giving an energy--driven outflow \citep{King2005ApJ,
  Zubovas2012ApJ}.

\subsection{Momentum-driven flows and $M-\sigma$ relation}

We estimate the cooling radius $R_{\rm C}$ by comparing the wind
cooling timescale \citep[cf.][]{King2003ApJ}:
\be
t_{\rm C} = \frac{2}{3} \frac{cR^2}{GM_{\rm BH}} \left(\frac{m_{\rm e}}{m_{\rm
    p}}\right)^2 \left(\frac{c}{v_{\rm w}}\right)^2 \simeq 10^7 R_{\rm kpc}^2
M_8^{-1} \; {\rm yr},
\ee
where $M_8$ is the SMBH mass $M_{\rm BH}$ in units of $10^8 \; \msun$, with
the ISM flow timescale
\be \label{eq:tflow}
t_{\rm flow} \sim \frac{R}{v_{\rm out,m}} \simeq 7 \times 10^6 R_{\rm kpc}
\sigma_{200} M_8^{-1/2} \left(\frac{f_{\rm g}}{f_{\rm c}}\right)^{-1/2} \; {\rm yr},
\ee
where $\sigma_{200}$ is the velocity dispersion in units of $200$~km/s,
  $v_{\rm out,m}$ is the {\em momentum--driven} outflow velocity
  \citep[see equation (14) in][]{King2003ApJ}, and $f_{\rm c} = 0.16$
  is the cosmological value of the baryon-to-dark-matter density
  fraction. Equating the two we find
\be \label{eq:rcool}
\begin{split}
R_C & \simeq \frac{3 GM}{2 c} \left(\frac{m_p}{m_e}\right)^2
\left(\frac{v_{\rm w}}{c}\right)^2 \left(\frac{f_{\rm g} \kappa \sigma^2}{2
  \pi G^2 M}\right)^{1/2} \\ & \sim 520 \; \sigma_{200}\; M_8^{1/2}\;
v_{0.1}^2 \left(\frac{f_{\rm g}}{f_{\rm c}}\right)^{1/2} \; \mathrm{pc}.
\end{split}
\ee 
Here $v_{0.1}$ is the wind velocity in units of $0.1c$. If we had
  used an
  expression for the energy--driven outflow velocity \citep[e.g. eq. 13
    from][]{King2011MNRAS} rather than the momentum-driven one, the cooling
radius would be smaller, so it is the momentum--driven velocity that is
relevant here. We note that the value of $R_{\rm C}$ depends strongly on the
assumed wind velocity; while theoretically we expect it to always be similar
to $0.1 c$, observations \citep{Tombesi2010A&A, Tombesi2010ApJ,
  Pounds2003MNRASb, Pounds2003MNRASa} suggest that it may vary from $\sim 0.03
c$ to $\sim 0.2 c$, leading to variations in $R_{\rm C}$ between $\sim 50$~pc
and $\sim 2$~kpc for a $10^8 \msun$ black hole in a $\sigma = 200$~km~s$^{-1}$
potential.

We see that close to the SMBH, outflows are momentum--driven. They only become
energy--driven once the SMBH reaches some critical mass which allows a driven
outflow formally to reach any radius \citep{King2005ApJ}. This mass is
\be 
\label{eq:msigma} M_\sigma =
\frac{f_{\rm g} \kappa}{\pi G^2} \sigma^4 \simeq 3.7 \cdot 10^8
\frac{f_{\rm g}}{f_{\rm c}} \sigma_{200}^4 \; \msun.  
\ee 
This value is very close to the observed $M-\sigma$ relation
\citep{Ferrarese2006ApJ}, provided that $f_{\rm g} \simeq f_{\rm c}$. Note
that in \citet{King2003ApJ}, the expression for $M_\sigma$ had an extra factor
of $1/2$ in it, because that paper simply assumed an escape velocity $\sim
\sigma$, rather than the full gravitational potential. \citet{King2005ApJ}
used a full isothermal potential, which gives the expression
(\ref{eq:msigma}).  In \citet{King2011MNRAS}, we mistakenly gave the incorrect
numerical value for $M_\sigma$, although we used the correct analytical
expression.

\subsection{Clearing the bulge of a spiral galaxy} \label{ref:spiralbulge}

Once the black hole mass reaches this value, the outflow can propagate to
large scales as the SMBH continues to grow. The timescale for the outflow
to reach the cooling radius is $\sim t_{\rm flow}$
(eq. \ref{eq:tflow}), i.e. $2-3$ Myr for the depleted bulge. Then the outflow
becomes energy driven and rapidly attains a much higher constant velocity
\be \label{eq:ve} 
 v_{\rm e} \simeq \left(\frac{2\eta f_{\rm g}'}{3f_{\rm
    c}}\sigma^2c\right)^{1/3} \simeq 925\sigma_{200}^{2/3}\left(\frac{f_{\rm
    g}'}{f_{\rm c}}\right)^{-1/3} \; {\rm km}\;{\rm s}^{-1}, 
\ee 
where $f_{\rm g}'$ is the gas fraction in the bulge during the outflow,
typically of the same order as $f_{\rm c}$ for young galaxies. This outflow
sweeps the bulge in an additional time
\be \label{eq:tout1} 
t_{\rm out, e} \simeq \frac{R_{\rm b} - R_{\rm C}}{v_{\rm e}} \sim (R_{\rm
  b,kpc} - 0.5) \; {\rm Myr}.
\ee
For typical bulge sizes of $1-2$~kpc, this amounts to another $\sim1$~Myr of
SMBH activity. Gas expelled from the bulge does not necessarily escape
  the galaxy altogether, but it may settle on to the galaxy disc, which is not
  affected by the outflow directly \citep{Nayakshin2012arXiv}, as part of the
  galactic fountain \citep{Bregman1980ApJ}.

Overall, this expulsion means that the SMBH must be active for $\sim 5$~Myr
after it reaches the $M - \sigma$ mass (\ref{eq:msigma}), even when the
  outflow propagates in an isothermal potential; in an NFW potential, the
  outflow would propagate faster and so the bulge would be cleared
  sooner. Even accreting at the Eddington rate (this may be possible
  contemporaneously with an outflow due to the presence of a massive accretion
  disc around the SMBH, which cannot be lifted by its wind due to large
  weight; see \citealt{Zubovas2012arXiv, Zubovas2011MNRAS,
    Nayakshin2010MNRAS}), the SMBH only grows its mass by $\sim10\%$ during
this time. Therefore, the value $M_\sigma$ (eq. \ref{eq:msigma}) is likely to
be the approximate upper limit to the mass of supermassive black holes in the
bulges of spiral galaxies.

\subsection{Elliptical galaxies and the transition to red-and-dead} \label{sec:reddead}

Following a major merger, a galaxy changes from spiral to elliptical
morphology. This means that the size of the spheroid grows from a bulge
several kpc across, to the scale of the whole galaxy, which may have a radius
of tens of kiloparsecs. Major mergers tend to cause bursts of star formation
followed by a burst of quasar activity some $\sim 300$~Myr later
\citep{Sanders1988ApJ,Gao1999ApJ,Canalizo2001ApJ,Combes2001sac}. This quasar
outburst is presumably what drives most of the remaining gas out of the galaxy
\citep{Zubovas2012ApJ}, transforming it into a red-and-dead elliptical.
Noting that the virial radius $R_{\rm V} \gg R_{\rm C}$, we may calculate
  the duration of quasar activity necessary to eventually drive the gas
  outside the virial radius of the galaxy \citep[cf.][]{King2011MNRAS}. Unlike
  in the simpler case of a bulge (eq. \ref{eq:tout1}), we have to consider the
  expansion of the outflow bubble even after the quasar has switched off. The
  result, in that case, is
\be \label{eq:tquas} 
t_{\rm q} \sim \frac{\sigma R_{\rm V}}{v_{\rm e}^2} \simeq \frac{1}{7H}
\left(\frac{\sigma}{v_{\rm e}}\right)^2 \sim 9 \times 10^7 \sigma_{200}^{2/3}
\left(\frac{f_{\rm g}}{f_{\rm c}}\right)^{2/3} \; {\rm yr} \simeq 2 t_{\rm
  Sal},
\ee 
where $H$ is the Hubble parameter and $t_{\rm Sal} = 45$~Myr is the
Salpeter time. We also allow some variation of the gas density by including
a dependence on $f_{\rm g}$. If the galaxy is almost devoid of gas already,
  the black hole only needs to be active for a very short time to drive an
  outflow past $R_{\rm V}$. This is also evident from the top panels of Figure
  \ref{fig:outflows}. For a high gas density $f_{\rm g} \sim f_{\rm
    c}$, as may be expected after a major gas-rich merger, we see that the
final mass of the black hole is
\be \label{eq:mfin}
M_{\rm fin} \sim M_\sigma \times {\rm exp}(2) \sim 7.5 M_\sigma.
\ee 

So the final mass of the SMBH in the centre of a red-and-dead elliptical
galaxy is expected to be almost an order of magnitude greater than that inside
a spiral galaxy with a red bulge exhibiting low star formation rate and
  low gas density.

This result disagrees with the estimate in \citet{Power2011MNRAS}, where we
found, using an energy argument, that the SMBH only needs to grow for
$\sim0.4t_{\rm Sal}$ in order to impart enough energy to the gas to let it
escape past the virial radius. There are two complications with this
calculation which give rise to the discrepancy. The first, minor, issue is
that we used the formal escape velocity from an isothermal potential, $v_{\rm
  esc} = 2 \sigma$, to obtain the required energy input. However, the escape
velocity is not well defined for an isothermal potential, and the adopted
value $2\sigma$ only moves gas to a position with radius a factor 20
larger. As a result, gas initially within $R_{\rm V} / 20 \sim 20$~kpc needs
more energy input to escape outside the virial radius. Nevertheless, this
effect by itself only produces a small correction.

The second issue is more important. As an escape condition, we compare the
virial radius of the galaxy with the radius of the contact discontinuity
between the wind and the outflowing ISM. Even while the quasar is active and
the outflow proceeds with velocity $v_{\rm e}$, two--thirds of the input
energy is transferred to the ISM {\it outside} the discontinuity
\citep{Zubovas2012ApJ}.  After the quasar switches off, this fraction
increases above 2/3. By the time the outflow stalls, numerical simulations
\citep{Zubovas2012arXiv} show that the energy content of the shocked
wind is only a few percent of the total input energy. This happens because
most of the shocked wind energy is thermal, and the wind cools as it expands,
transferring energy to the outflow; by the time the outflow stalls, the energy
per unit mass is roughly the same in both the wind and the ISM, so the ISM,
which contains most of the mass, also has most of the energy.

All this means that in order to get the inner surface of the swept--up host
ISM (the contact discontinuity) to expand past $R_{\rm V}$, the quasar has to
be active for at least a few times longer, and perhaps an order of magnitude
longer, than estimated by \citet{Power2011MNRAS}. This correction brings the
quasar duration required by energy arguments \citet{Power2011MNRAS} into
agreement with that required by the bubble dynamics, as here.

We can now combine the final SMBH mass derived above (eq. \ref{eq:mfin}) and the
bulge stellar mass estimate from \citet{Power2011MNRAS},
\be \label{eq:mbulge}
M_{\rm b} \lesssim 0.6 \zeta M_{\rm g, vir} \sigma_{200},
\ee
where $\zeta \sim 1$ is a factor encompassing our uncertainty regarding the
amount of feedback that self-regulated star formation produces when compared
with the typical gas momentum in the bulge, and $M_{\rm g,vir}$ is the total
gas mass inside the virial radius of the galaxy. Together with the formulae
for mass of an isothermal sphere and the virial radius (equations 3 \& 18 in
\citet{King2011MNRAS}, respectively), they give us the black hole - bulge mass
relation:
\begin{equation}
\frac{M_{\rm fin}}{M_{\rm b}} \simeq 7.5 \times 3.7 \times 10^8 \sigma_{200}^{4}
\frac{7 G H}{0.6 \zeta \times 2 f_{\rm c} \sigma^3 \sigma_{200}} \sim 3 \times
10^{-3} \zeta^{-1}.
\end{equation}
This is consistent with the observed relation
\citep[e.g.][]{Haering2004ApJ}. For spiral galaxies, the black hole mass
  is approximately an order of magnitude lower, but the spheroidal component
  is also smaller by a similar factor, so the relation holds roughly in this
  case as well.

\section{Galaxy environment and its influence on the $M-\sigma$
  relation} 
\label{sec:clusters}

So far, in our calculations we have assumed that $f_{\rm g} \simeq f_{\rm c}$
throughout the evolution of a galaxy. This may be a reasonable assumption on
scales larger than $R_{\rm V}$, but is unlikely to be correct on the smaller
scales that most immediately affect the properties of the black hole. Galaxy
bulges are baryon-dominated and so by the time the SMBH begins
driving large--scale outflows, the fraction of gas remaining in the bulge
depends mainly on depletion by star formation and refilling via mergers or
accretion of cold gas from the halo.

\subsection{Gas depletion}

When a galaxy forms, its central regions have $f_{\rm g} \sim 1$, as most
  of the baryons have not yet had time to form stars. Subsequently, star
  formation depletes the gas on a timescale $t_{\rm SF}$. As a rough
  approximation, we assume that star formation persists throughout the
  lifetime of the galaxy, with $2\%$ of gas converted into stars on a
  dynamical time \citep{Kennicutt1998ApJ}. With a scale radius for the bulge
  of $1$~kpc, typical for a spiral galaxy, this gives a star formation
  timescale
\begin{equation}
t_{\rm SF} = \frac{t_{\rm dyn}}{\epsilon_{\rm SF}} \simeq 2.5 \times
10^8 R_{\rm kpc} \sigma_{200}^{-1} \epsilon_{0.02}^{-1} \; {\rm yr}
\simeq 5.5 t_{\rm Sal}.
\end{equation}
If the SMBH grows from a seed mass of $\sim 10 \msun$, it takes $\sim 16
t_{\rm Sal}$ to reach $10^8 \msun$, by which time the gas fraction is depleted
to
\begin{equation}
f_{\rm g}' \simeq {\rm exp}\left(-\frac{16 t_{\rm Sal}}{t_{\rm SF}}\right)
\sim 0.05 \simeq 0.35 f_{\rm c}.
\end{equation}
Substituting $f_{\rm g}'$ instead of $f_{\rm g} = f_{\rm c}$ into
  equation \ref{eq:msigma} we find that in spiral galaxies we may expect SMBH
  masses to be a factor $\sim 2.5$ lower than predicted above.

The estimate above, however, is highly uncertain. At least three major
  effects may drive it to either higher or lower estimates for the SMBH
  mass. The estimate may be decreased if the SMBH growth is intermittent,
  i.e. the duty cycle is low, while star formation efficiency is approximately
  constant throughout the lifetime of the galaxy. In that case, the gas
  fraction in the bulge may be depleted to even lower values than $f_{\rm
    g}'$, leading to a smaller critical SMBH mass. If, for example, the black
  hole duty cycle is $10\%$ \citep{Schawinski2007MNRAS}, then the critical
  black hole mass drops to a very low value of $10^4 \; \msun$. The fact that
  such small black holes are not observed suggests that star formation and
  SMBH growth are temporally correlated in galaxies.

On the other hand, there have been suggestions that SMBHs may grow from
  seeds as massive as $10^5 \; \msun$ \citep[e.g.][]{Begelman2006MNRAS}. In
  this case, the SMBH growth timescale is shorter than the $16 t_{\rm Sal}$
  used above, and so the gas is only depleted to $f_{\rm g} \sim 0.4$,
  requiring a larger final SMBH mass.

Finally, a third complication may affect the slope of the predicted
  $M_{\rm BH} - \sigma$ relation. Galaxies with larger $\sigma$ have
  larger effective radii and larger bulges \citep[e.g.][]{Marconi2003ApJ} and
  so presumably are less efficient in forming stars. As a result, the critical
  value of $f_{\rm g}$ may correlate positively with $\sigma$, increasing the
  slope of the mass - velocity dispersion relation. This may be important when
  comparing spiral and elliptical galaxies: not only do elliptical galaxies
  need to grow their SMBHs for longer in order to expel the gas past the
  virial radius (see Section \ref{sec:reddead}), they must do so while the
  galaxy has a higher gas fraction than a spiral galaxy would. However, we
  cannot quantify this effect in more detail without recourse to a much more
  detailed model of long-term galaxy evolution, which is beyond the scope of
  this paper.

A conclusion of this section is that it is likely, but by no means
  certain, that isolated spiral galaxies could have central SMBHs that are a
  factor $\sim 2.5$ less massive than predicted previously. Isolated
  elliptical galaxies may be similarly offset from the relation given in
  eq. (\ref{eq:mfin}). In addition, the slope in the $M - \sigma$ relation for
  these galaxies may be slightly steeper than $M \propto \sigma^4$ due to a
  correlation between the velocity dispersion and size of the galaxy
  spheroidal components.

\subsection{Gas replenishment in clusters}

The calculations above are valid for field galaxies, where internal
  processes of gas depletion dominate. If a galaxy falls inside a cluster, ram
  pressure and tidal forces strip most of its gas away
  \citep[e.g.][]{Takeda1984MNRAS}. This process lowers the average gas
  fraction in the galaxy; however, since gas is predominantly stripped from
  the outskirts of the galaxy \citep{Abadi1999MNRAS} and star formation is
  suppressed in line with gas depletion \citep{Lewis2002MNRAS,Balogh2000ApJ},
  the limiting central SMBH mass is not significantly affected by this
  process.

The cluster environment may also provide a replenishment of gas for
  galaxies that do not experience significant stripping, such as Brightest
  Cluster Galaxies (BCGs). These galaxies sit close to the centre of the
  cluster potential and thus are embedded in a massive hot halo. Cooling flows
  from the halo replenish gas in the galaxy \citep{Fabian1994ARA&A},
  sustaining star formation and feeding the SMBH. If this replenishment
  happens on a timescale which is not too different from the cluster dynamical
  timescale $t_{\rm d, cl} = R_{\rm cl} / \sigma_{\rm cl}$, then the two
  effects of star formation and cooling flow replenishment balance each other
  at a gas fraction
\begin{equation}
f_{\rm eq} = \frac{t_{\rm SF}}{t_{\rm d,cl} + t_{\rm SF}}.
\end{equation}
Using the same values as above for star formation and $R_{\rm cl} = 1$~Mpc and
$\sigma_{\rm cl} = 1000$~km/s, we find
\begin{equation}
f_{\rm eq} = 0.2 \simeq f_{\rm c}.
\end{equation}
We see that BCGs, and perhaps some other cluster galaxies which are
  moving slowly with respect to the cluster halo gas, should have black hole
  masses governed by the $M_\sigma$ of equations \ref{eq:mfin} and
  \ref{eq:msigma}. The latter equation may apply to the strongly perturbed
  spiral galaxies which have lost most of their disc and so transformed to S0
  morphological type, but still have only a small bulge which can be cleared
  easily and rapidly by the AGN.

\section{Discussion} \label{sec:discuss}

\begin{figure}
  \centering
  \includegraphics[width=0.45\textwidth]{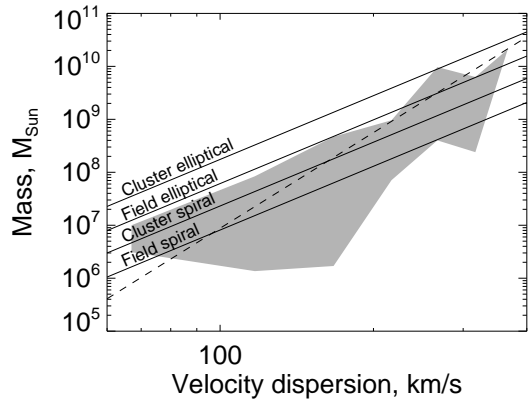}
  \caption{The four $M-\sigma$ relations (solid lines) and their combined
    effect on observational fits (dashed line). All solid lines have slopes $M
    \propto \sigma^4$ and the dashed line has $M \propto \sigma^6$. The grey
    area is the approximate locus of data points in Fig. 3 of
    \citet{McConnell2011Natur}.}
  \label{fig:msigdiag}
\end{figure}

We have shown that supermassive black holes residing in the bulges of spiral
galaxies should have an $M-\sigma$ relation that can be approximated by a
power law slightly steeper than $M \propto \sigma^4$, if gas depletion
due to star formation is considered. Typically, such galaxies would have $M <
M_\sigma$ as predicted in our earlier papers
\citep{King2003ApJ,King2010MNRASa,King2011MNRAS}. During an event triggering
strong central accretion, such as a major merger, the SMBH is likely to grow
by a further factor of $\lesssim 7.5$ (i.e. for $\lesssim 2$ Salpeter times)
before it can drive the gas out of the galaxy halo. This factor is an upper
limit, because we calculated it for an isothermal background potential, which
is the most difficult to escape from.

One might think that since the SMBH is not immediately aware of the processes
occurring around the virial radius of the halo, it could grow for longer than
the calculated $\sim 2$ Salpeter times, as long as the gas reservoir feeding
it is not depleted. However, the total mass of gas that must be accreted in
this process is of order $10^9 \msun$, perhaps even more. This is too high to
reside in a single accretion disc, and the SMBH is very probably fed in a
series of stochastic feeding events \citep{King2007MNRAS,
  Nayakshin2007arXiv}. Feedback therefore affects the SMBH feeding reservoir,
and so it seems more likely that the reservoir is fairly efficiently
depleted. While gas may fall back on to the SMBH if it has not yet escaped the
halo and trigger subsequent bursts of activity, it cannot do so once the
halo is depleted, so the AGN should switch off very soon after the outflow
clears the galaxy.

Our results suggest that there should be a difference between the $M - \sigma$
relations for spiral star-forming galaxies and for red and dead
ellipticals. The latter should have SMBHs that are systematically overmassive
when compared with the prediction of the $M - \sigma$ relation for the spiral
galaxies. This mass difference is not huge - less than an order of magnitude -
but it may become apparent when better SMBH mass and spheroid velocity
dispersion measurements become available.

There is another possible offset in the $M - \sigma$ relation depending on the
galaxy environment. Galaxies in clusters, especially those residing near
  cluster centres, may accrete gas from the hot cluster halo via cooling
  flows. This process may balance the depletion by star formation at a gas
fraction $\sim 0.2$, preserving the old $M - \sigma$ relationship
(eq. \ref{eq:msigma}) in galaxies that had been spirals by the time they
  entered the cluster environment; even though they lose most of the gas from
  the outskirts, the central spheroidal component remains unperturbed and the
  analysis in Section \ref{ref:spiralbulge} still applies. Cluster
ellipticals would also be a factor $\lesssim 7.5$ more massive. On average,
cluster galaxies have black holes slightly more massive (by a factor $\sim
2.5$) than their field counterparts. Our findings are consistent with the
  recent discovery of two SMBHs in the centres of giant elliptical galaxies
  that have masses in excess of $10^{10} \; \msun$. They are both Brightest
  Cluster Galaxies and our model predicts that their masses should be
  approximately an order of magnitude above the relation derived in earlier
  works.

In summary, we have found that there may be three (or possibly even
four) $M-\sigma$ relations in total, depending on galaxy morphology
and environment:

\begin{itemize}
\item spiral galaxies with evolved bulges have $M_{\rm BH} \sim 1.3 \times
  10^8 \sigma_{200}^{4+\beta} \msun$ with $\beta \lesssim 1$ depending on star
  formation efficiency and the relation between the bulge size and $\sigma$;
\item spiral galaxies with evolved bulges residing in gas-rich cluster
  environments have $M_{\rm BH} \sim 3.7 \times 10^8 \sigma_{200}^4 \msun$
  (the original $M-\sigma$ relation); such galaxies, however, may be extremely
  rare due to high merger probability in clusters;
\item field elliptical galaxies have $M_{\rm BH} \sim 9.8 \times 10^8
  \sigma_{200}^{4+\zeta} \msun$;
\item elliptical galaxies close to cluster centres have $M_{\rm BH} \sim 28
  \times 10^8 \sigma_{200}^4 \msun$.
\end{itemize}

There is some uncertainty in the slope, especially for field galaxies. It is
clear that, as bigger bulges have larger velocity dispersions, the slope
should be steeper than $4$, but its precise value depends strongly on the
assumed relation between $R_{\rm b}$ and $\sigma$ (both its slope and
intercept).

The differences among the four relations are small, at most factors of a few
in the normalization. Intrinsic scatter in each population may blur the
distinctions further. In addition, since elliptical galaxies tend to have
higher velocity dispersions than spirals, and cluster galaxies than field
galaxies, the overall effect of combining the relations may be to increase the
observed slope of the $M-\sigma$ relation. We illustrate this in Figure
\ref{fig:msigdiag}. We plot the four $M-\sigma$ relations (assuming $\beta =
0$) over a polygon representing the approximate locus of data points taken
from Figure 3 of \citet{McConnell2011Natur}. All four lines intersect the
locus. The large fraction of the locus being below the four lines is
reasonable since the predicted relations are only upper limits for
  different environments. Furthermore, there is a hint that higher values of
$\sigma$ correspond to higher $M-\sigma$ relations. We also plot a line with
$M \propto \sigma^6$, which roughly bisects the observed locus and so might
emerge from a fit to the data unsorted by galaxy type, especially in the
  $\sigma > 100$~km/s part, where we do not expect contamination from nuclear
  stellar clusters \citep{McLaughlin2006ApJ, Ferrarese2006ApJ}.

There have been recent tentative claims that the $M-\sigma$ relation may be
slightly different for different galaxy types, and that the general relation
may show an upturn at higher values of $\sigma$
\citep{Marconi2003ApJ,McConnell2011Natur}. Our considerations here suggest
that both features may be natural consequences of wind feedback.

\section*{Acknowledgments}

We thank the referee for very detailed and thorough comments which allowed us
to significantly improve the clarity of the paper.

Research in theoretical astrophysics at Leicester is supported by an STFC
Rolling Grant. KZ is supported by an STFC research studentship.


\end{document}